\documentclass[aps,pre,twocolumn,groupedaddress]{revtex4}

\newif\ifpdf
\ifx\pdfoutput\undefined
\pdffalse 
\else
\pdfoutput=1 
\pdftrue
\fi

\ifpdf
\usepackage[pdftex]{graphicx}
\else
\usepackage{graphicx}
\fi
\usepackage{hyperref}

\begin{document}

\ifpdf
\DeclareGraphicsExtensions{.pdf, .jpg}
\else
\DeclareGraphicsExtensions{.eps, .jpg}
\fi

\def\hslash{\hbar}
\def\imag{i}
\def\grad{\vec{\nabla}}
\def\div{\vec{\nabla}\cdot}
\def\curl{\vec{\nabla}\times}
\def\DDt{\frac{d}{dt}}
\def\ddt{\frac{\partial}{\partial t}}
\def\ddx{\frac{\partial}{\partial x}}
\def\ddy{\frac{\partial}{\partial y}}
\def\lap{\nabla^{2}}
\def\divv{\vec{\nabla}\cdot\vec{v}}
\def\gradS{\vec{\nabla}S}
\def\vvec{\vec{v}}
\def\wc{\omega_{c}}
\def\<{\langle}
\def\>{\rangle}
\def\Tr{{\rm Tr}}
\def\Csch{{\rm csch}}
\def\Coth{{\rm coth}}
\def\Tanh{{\rm tanh}}
\def\g2{g^{(2)}}
\newcommand{\al}{\alpha}

\newcommand{\la}{\lambda}
\newcommand{\del}{\delta}
\newcommand{\om}{\omega}
\newcommand{\ep}{\epsilon}
\newcommand{\pd}{\partial}
\newcommand{\bra}{\langle}
\newcommand{\ket}{\rangle}
\newcommand{\bbra}{\langle \langle}
\newcommand{\kket}{\rangle \rangle}
\newcommand{\non}{\nonumber}

\title{Quantum transport in chains with noisy off-diagonal couplings}

\author{Andrey Pereverzev}
\affiliation{Department of Chemistry and Center for Materials Chemistry, 
University of Houston \\ Houston, TX 77204}
\author{Eric R. Bittner}
\affiliation{Department of Chemistry and Center for Materials Chemistry, 
University of Houston \\ Houston, TX 77204}
\date{\today}

\begin{abstract}
We present a model for conductivity and energy diffusion in a linear
chain described by a quadratic Hamiltonian with Gaussian noise.  We
show that when the correlation matrix is diagonal, the noise-averaged
Liouville-von Neumann equation governing the time-evolution of the
system reduces to the Lindblad equation with Hermitian Lindblad
operators.  We show that the noise-averaged density matrix for the
system expectation values of the energy density and the number density
satisfy discrete versions of the heat and diffusion
equations. Transport coefficients  are given
in terms of model Hamiltonian parameters.  We discuss conditions on the Hamiltonian 
under which 
the noise-averaged expectation value of the total
energy remains constant.  For chains placed between two heat
reservoirs, the gradient of the energy density along the chain is
linear.

 \end{abstract}


\maketitle
\section{Introduction}
The problem of energy and charge transport in low-dimensional systems has
attracted considerable current attention. 
\cite{buldum,saito,macedo:155309,kolovsky:046202,woljcik:230602,richter:206801}.
This recent interest is primarily driven by new advances in nanotechnology
and the resulting need to explain new and unusual transport properties
of nanomaterials \cite{schwab,pouthier}. In many respects, such
materials behave as effectively one or two dimensional systems. 
Furthermore, from a theoretical standpoint, this tnterest is driven by the desire to understand
how dimensionality affects transport properties
and  whether normal transport
is possible in low dimensions, and, if so, what are the conditions for
such behavior. 

Studies of heat transport in one dimensional models have long and rich
history.  Review articles on heat transport in one dimension include
Refs. \cite{bonetto:128} and \cite{lepri:1}.  The earliest studies
addressing this problem for anharmonic chains include the seminal
numerical simulations by Fermi, Pasta and Ulam \cite{fermi:1} and
Peierls' discussion of the constraints on phonon interactions in one
dimension \cite{Peierls}.  More recent theoretical studies of heat transport
in one dimension include Refs. \cite{lepri:2, lepri:1,narayan:1, li:1, 
pereverzev:1}.
Extensive numerical simulations have also
been reported in recent years \cite{li:2, li:3, li:4, lepri:1, lepri:067102}.  
It is now well established that thermal transport in momentum conserving
one dimensional models is anomalous, i.e. the Fourier heat law is not
observed in this systems. The physical reason for such behavior is in the
slow decay of normal modes with 
small wave numbers. In one dimension,
the fast propagation of heat carried by these modes makes normal
heat transport impossible.

Due to the
difficulties in the theoretical treatment of one-dimensional
transport, exactly solvable models are of interest. One of the few one dimensional
models that shows normal heat conductivity is a stochastic and non-Hamiltonian model
introduced by Bolsterli {\it et al.} \cite{bolsterli:1086, bonetto:2}. 
In this paper we consider a quantum mechanical Hamiltonian model in which the 
Hamiltonian matrix elements are taken as time dependent stochastic 
variables.  We show that upon averaging over the noise, one obtains 
a Lindblad-type equation for the evolution of the quantum density matrix. 
We then go on to give the conditions upon the Hamiltonian
which produce normal (i.e. diffusive) energy and particle transport.
Finally, we demonstrate the validity of our model 
using numerical simulations. 

\section{Gaussian white noise Hamiltonians} \label{2}
The Hamiltonian of the one-dimensional model that we will consider in
this paper belongs to a general class of Hamiltonians with Gaussian
white noise
\begin{eqnarray}
H=H_0+\sum_i \la_i(t) V_i, \label{noisy}
\end{eqnarray}
where $H_0$ and $V_i$ are arbitrary Hermitian operators and gaussian
stochastic coefficients $\la_i(t)$ have average mean
$\overline{\la_i(t)}  = 0$ and second moments
given by
\begin{eqnarray}
\overline{\la_i(t)\la_j(t')}=g_{ij}\delta(t-t'). \label{second}
\end{eqnarray}
Hamiltonians such as this can model a wide variety of physical
situations where the motion or transport is driven by an external
field.  One such example is the case of F{\"o}rster resonant
excitation transfer  in biomolecules where the migration and
diffusion of an initial electronic excitation within the system is
dependent upon the local environmental and conformational fluctuations
that modulate the off-diagonal couplings. 

From a theoretical standpoint,  such Hamiltonians are useful since
one can explicitly
average over the noise when calculating the time evolution of the
density matrix or expectation values of various observables.
Noise-averaged time evolution for various specific forms of the
Hamiltonian in Eq. (\ref{noisy}) have been considered previously
\cite{ovchinnikov:733, madrukar1424, jayannavar:553, girvin:4896,
fischer:1578}.  We give here a brief discussion of noise-averaged time
evolution for the general case and then go on to emphasize points that
are relevant for the studies of heat transport.

In general, the quantum density matrix satisfies the Liouville-von
 Neumann equation,
\begin{eqnarray}
i\frac{\partial\rho}{\partial t}=({\cal L}_0+{\cal L}_V(t))\rho.
\end{eqnarray}
Action of the superoperators ${\cal L}_0$ and ${\cal L}_V$ on $\rho$
is given, respectively, by ${\cal L}_0\rho=\frac{1}{\hbar}\left[H_0,\rho\right]$
and ${\cal L}_V(t)\rho=\frac{1}{\hbar}\sum_i\la_i(t)\left[V_i,\rho\right]$.  The
density matrix at time $t$ is given in terms of the density matrix
$\rho(0)$ at time $t=0$ by
\begin{eqnarray}
\rho(t)={\cal U}(t)\rho(0).
\end{eqnarray} 
Here the time-evolution superoperator ${\cal U}(t)$ is given by the
following infinite series:
\begin{widetext}
\begin{equation}
{\cal U}(t )=e^{-i {\cal L}_0t}-i\int_0^t d\tau e^{-i{\cal L}_0(t-
\tau)}{\cal L}_V(\tau)e^{-i{\cal L}_0\tau} \nonumber  -\int_0^t
d\tau \int_0^\tau d\tau'e^{-i{\cal L}_0(t-\tau)} {\cal L}_V(\tau)
e^{-i{\cal L}_0(\tau -\tau ')}{\cal L}_V(\tau')e^{-i{\cal L}_0\tau'}+...
\label {series}
\end{equation}
\end{widetext}
Noise-averaged expectation  values of an operator $O$
are computed using
\begin{eqnarray} 
\overline{\bra O(t)\ket}={\rm Tr}\left(\overline{{O(t){\cal
U}(t)\rho(0)}}\right). \label{aver}
\end{eqnarray}
It is assumed here that operator $O$ can have an explicit time
dependence.  When performing averages as in Eq. (\ref{aver}), we need to
distinguish between two types of operators: those with and those without
stochastic coefficients $\la_i(t)$.  In the latter case, averaging
over noise as in Eq. \ref{aver} reduces to an averaging of the
evolution superoperator ${\cal U}(t)$ and such expectation values can
be calculated with the noise averaged density matrix.

Noise averaging of ${\cal U}(t)$ can be performed by taking averages for each
term in the series and then resumming the series. This involves
averaging products of the stochastic coefficients.  Since
$\lambda_i(t)$ is sampled from a Gaussian deviate, all terms involving
an odd number of coefficients necessarily vanish.  Furthermore, any
term with an even number of terms can be written as a sum of all
possible products of second moments. However, due to the order of
integrations over time in Eq. (\ref{series}) and the fact that second
moments in Eqs. (\ref{second}) involve delta functions in time, only one
product from the sum contributes to the average after all the time
integrations are performed.  For example,  the fourth order
average
$\overline{\la_i(t)\la_j(\tau)\la_k(\tau')\la_l(\tau'')}$,
decomposes into the following second-order terms:
\begin{eqnarray}
& &\overline{\la_i(t)\la_j(\tau)\la_k(\tau')\la_l(\tau'')} 
=g_{ij}g_{kl}\delta(t-\tau)\delta(\tau'-\tau'') \non \\
& & +g_{ik}g_{jl}\delta(t-\tau') \delta(\tau-\tau'')
+g_{il}g_{jk}\delta(t-\tau'')\delta(\tau-\tau')  \non \\
\end{eqnarray}
Since the region of integration for the fourth order term in the
series (\ref{series}) is $t\geq \tau\geq\tau' \geq \tau''$ we can
see that only the $g_{ij}g_{kl}\delta(t-\tau)\delta(\tau'-\tau'') $
term will contribute to the series. Similar analysis can be applied to
all even order terms.
 
Averaging over noise and resumming the series given by Eq. (\ref{series})
produces
\begin{eqnarray}
\overline{{\cal U}(t)}=e^{-i{\cal L}_0t-{\cal M}t}, \label{aveprop}
\end{eqnarray}
where the action of the superoperator ${\cal M}$ on the density matrix
$\rho$ is given by
\begin{eqnarray}
{\cal M}\rho=\frac{1}{2\hbar^2}
\sum_{ij}g_{ij}\left[V_i,\left[V_j,\rho\right]\right]. \label{M}
\end{eqnarray}
Hence, our noise-averaged propagation is unitary and norm-conserving. 
It follows from Eq. (\ref{aveprop})  that the noise-averaged density
matrix satisfies the following equation
\begin{eqnarray}
i\frac{\partial \overline{\rho}}{\partial t} =({\cal L}_0-i{\cal
M})\overline{\rho}. \label{avedens}
\end{eqnarray}
This allows one to see an interesting connection between
 noise-averaged time evolution of the density matrix for the noisy
 system and the time evolution of the reduced density matrix for an
 open quantum system.

Consider the case when correlation matrix $g_{ij}$ in Eq. (\ref{M}) is
 diagonal, i.e.  $g_{ij}=g_{ii}\delta_{ij}$.  In this case,
 Eq. (\ref{avedens}) can be rewritten as
\begin{eqnarray}
i\frac{\partial \overline{\rho}}{\partial t} =
\frac{1}{\hbar}\left[H_0,\overline{\rho}\right]
-\frac{i}{2\hbar^2}\sum_i
g_{ii}\left[V_i,\left[V_i,\overline{\rho}\right]\right].
\end{eqnarray}
This has the form of a Lindblad equation with Hermitian Lindblad
operators  \cite{lindblad}. Thus, the time evolution of the reduced
density matrix under the Lindblad equation with Hermitian operators is
equivalent to noise-averaged unitary evolution with Hamiltonian
(\ref{noisy}) in the special case of a diagonal correlation matrix
$g_{ij}$.  Connection between Lindblad-type evolution with Hermitian
Lindblad operators and an averaged random unitary evolution was
discussed in Ref. \cite{salgado:975}.  It is interesting that the
unitary operator proposed in Ref. \cite{salgado:975} is different from
the unitary operator corresponding to the Hamiltonian in
Eq. (\ref{noisy}).  This implies that averaging different unitary
evolutions over the same noise spectrum can ultimately lead to the
same Lindblad-type evolution.

If the operator whose noise-averaged expectation value we want to find
depends on the coefficients $\la_i(t)$ then these coefficients must be
taken into account when performing averages as in Eq. \ref{aver}.
Let us consider the linear case $\la_i(t)W$ where $W$ is
time independent and $\lambda_i(t)$ is a gaussian random variable. 
 An example of an operator of this type is the
second term of Eq. \ref{noisy}.  Using series (\ref{series}) for
superoperator ${\cal U}(t)$ and substituting $\la_i(t)W$ for $O$ in
Eq.(\ref{aver}) we can again perform averages for each term and resum
the series to obtain
\begin{widetext}
\begin{eqnarray}
{\rm Tr}\left(\overline{\la_i(t)W{\cal
U}(t)\rho(0)}\right)=-\frac{i}{\hbar}{\rm
Tr}\left(W\sum_{j}g_{ij}\left[V_j,e^{-i{\cal L}_0t-{\cal M}t}\rho(0)\right]
\right) =-\frac{i}{\hbar}{\rm
Tr}\left(\overline{\rho(t)}\sum_{j}g_{ij}\left[W,V_j\right]\right)\label{ave2}
\end{eqnarray}
where the last line is obtained by using cyclic permutations under the
trace and the fact that $e^{-i{\cal L}_0t-{\cal
M}t}\rho(0)=\overline{\rho(t)}$.
\end{widetext}

Explicit averaging over noise was performed in Eqs. (\ref{aveprop}) and
(\ref{ave2}).  However, these expressions remain mostly formal. For some
forms of operators $H_0$ and $V_i$ further exact treatment is
possible. One such example is $H_0$ and $V_i$ that are quadratic forms
of creations and annihilation operators.  Hamiltonians of this type
were used in solid state physics to describe electronic transport
under the influence of an external driving field such as lattice
vibrations \cite{girvin:4896,haug}. These models gave a satisfactory
picture of electron diffusion.  However, such models generally do
not conserve average energy of the system and therefore cannot show
normal heat propagation.

Let us use Eqs. (\ref{aveprop}) and (\ref{ave2}) to determine the precise
conditions that the operators $H_0$ and $V_i$ must satisfy in order
for the noise-averaged expectation value of the total energy to remain
constant.  The noise-averaged energy expectation value is given by
\begin{eqnarray}
\overline{\bra H(t)\ket}=\overline{\bra H_0(t)\ket}+\sum_i
\overline{\la_i(t)\bra V_i(t)\ket} \label{aveene}
\end{eqnarray}
Since $H_0$ does not depend upon  the stochastic coefficients,  we obtain
\begin{eqnarray}
\overline{\bra H_0(t)\ket}={\rm Tr}( H_0 e^{-i{\cal L}_0t-{\cal
M}t}\rho(0)).
\end{eqnarray}
To find the second term on the right hand side of Eq. (\ref{aveene}) we
use Eq. (\ref{ave2}) where we need to put $V_i$ for $W$ and sum over all
$i$.  We obtain
\begin{eqnarray}
\sum_i \overline{\la_i(t)\bra V_i(t)\ket}= -\frac{i}{\hbar}{\rm
Tr}\left(\overline{\rho(t)}\sum_{ij}g_{ij}\left[V_i,V_j\right]\right)
\end{eqnarray}
Since matrix $g_{ij}$ is always symmetric, the last quantity vanishes
for arbitrary $V_i$'s and, therefore, does not contribute to
$\overline{\bra H(t)\ket}$. Thus we have
\begin{eqnarray}
\overline{\bra H(t)\ket}=\overline{\bra H_0(t)\ket}
\end{eqnarray}
For $\overline{\bra H(t)\ket}$ to be constant in time, its time
derivative must vanish.  Differentiating $\overline{\bra H_0(t)\ket }$
with respect to time we obtain
\begin{eqnarray}
\frac{d \overline{\bra H_0(t)\ket }}{d t}={\rm Tr}( H_0( -i{\cal
L}_0-{\cal M})\overline{\rho(t)}).
\end{eqnarray}
Using the fact that the trace is invariant under cyclic permutations,
the symmetric nature of matrix $g_{ij}$, and that ${\cal L}_0H_0=0$,
we obtain
\begin{eqnarray} 
\frac{d \overline{\bra H_0(t)\ket }}{d t}=-{\rm
Tr}\left(\overline{\rho(t)}{\cal M}H_0\right).
\label{deriv}
\end{eqnarray}
For arbitrary $\rho(0)$
this derivative will be equal to zero if 
\begin{eqnarray}
{\cal M}H_0=\frac{1}{2\hbar^2}
\sum_{ij}g_{ij}\left[V_i,\left[V_j,H_0\right]\right]=0. \label{zero}
\end{eqnarray}
This is the condition for $\overline{\bra H(t)\ket}$ to be constant in
time. Examples  of operators $H_0$ and $V_i$ for which 
 Eq. (\ref{zero}) is true include the case
  when $H_0$ and $V_i$  satisfy the condition $[H_0,V_i] = const$ for all $i$,
in particular when $H_0$ and $V_i$ commute.

\section{Model Hamiltonian and observables}
We now apply the results of the previous section to study quantum heat
 conduction and diffusion in a one-dimensional chain of $N$ sites
 described by the following Hamiltonian
\begin{eqnarray}
H&=&\hbar \om \sum_{i}^{N}a_i^{\dagger}a_i +
\hbar\sum_{i=1}^N\mu_i(t)(a_i^{\dagger}a_{i+1}+ a_{i+1}^{\dagger}a_i)
\non \\ & &+ i\hbar\sum_{i=1}^N\nu_i(t)(a_i^{\dagger}a_{i+1}-
a_{i+1}^{\dagger}a_i)\label{Ham}
\end{eqnarray}
Here $a_i^{\dagger}$ and $a_i$ denote, respectively, creation and
annihilation operators for quasi-particles at chain site $i$.
These quasi-particles can be bosons, fermions, or spin excitations.
The stochastic hopping coefficients $\mu_i(t)$ and $\nu_i(t)$ are
 realizations of the Gaussian noises with zero mean. We choose the
 following simple expressions for the second moments,
\begin{eqnarray}
\overline{\mu_j(t)\mu_k(t')}&=&\alpha^2_1\del_{jk}\del(t-t'), \non \\
 \overline{\nu_j(t)\nu_k(t')}&=&\alpha^2_2\del_{jk}\del(t-t') \non \\
 \overline{\mu_j(t)\nu_k(t')}&=&0
\label{corr}
\end{eqnarray}
We will consider two types of boundary conditions: cyclic and open
ends.  For the former we have $a_{N+1}=a_1$,
$a^{\dagger}_{N+1}=a^{\dagger}_1$, and for the latter
$a_{N+1}=a^{\dagger}_{N+1}=0$.

The model Hamiltonian in Eq. (\ref{Ham}) belongs to the class of noisy
Hamiltonians we just described.  Indeed, we can rewrite Eq. \ref{Ham}
as
\begin{eqnarray}
H=H_0+\sum_{j=1}^{2N}\la_j(t)V_j \label{Ham2}
\end{eqnarray}
where 
$$H_0=\hbar \om \sum_{i}^{N}a_i^{\dagger}a_i$$ and 
$$\la_j(t)V_j=\hbar\mu_i(t)(a_i^{\dagger}a_{i+1}+
a_{i+1}^{\dagger}a_i)$$ for even $j$'s and 
$$\la_j(t)V_j=i\hbar\nu_i(t)(a_i^{\dagger}a_{i+1}-
a_{i+1}^{\dagger}a_i)$$ for odd $j$'s.  Thus, discussion of
Sec. \ref{2} can be applied to the studies of time evolution driven by
this Hamiltonian. Note that Hamiltonian (\ref{Ham}) satisfies the
average energy constraint in Eq. (\ref{zero}) because its time
independent part $H_0$ commutes with all $V_i$'s.  Therefore noise
averged expectation value of the total energy remains constant in time
and we can discuss heat conduction in this model.

Heat conduction and particle diffusion can be studied by considering
time evolution of expectation values of energy density and number
density averaged over noise.  For both types of boundary conditions,
operators for the number density $n_i$ are
\begin{eqnarray}
n_i = a_i^{\dagger}a_i.
\end{eqnarray}
The energy density $\epsilon_i$ at site $i$ for cyclic boundary
conditions is given
\begin{eqnarray}
\epsilon_i&=&\hbar \om a_i^{\dagger}a_i +\frac{1}{2}\hbar\mu_i(t)A_i
+\frac{1}{2}\hbar\mu_{i-1}(t)A_{i-1} \non \\ &
&+\frac{1}{2}\hbar\nu_i(t)B_i +\frac{1}{2}\hbar\nu_{i-1}(t)B_{i-1}
\label{observables}
\end{eqnarray}
where $A_i=(a_{i}^{\dagger}a_{i+1}+a_{i+1}^{\dagger}a_{i})$ and
$B_i=i(a_{i}^{\dagger}a_{i+1}-a_{i+1}^{\dagger}a_{i} )$ In the case of
open-end boundaries, operators for the energy density at site $i$ are
the same as for cyclic boundaries except for the first and last sites
where we use
\begin{eqnarray}
\epsilon_{1}&=&\hbar \om a_{1}^{\dagger}a_{1} 
+\frac{1}{2}\hbar\mu_{1}(t)A_{1} +\frac{1}{2}\hbar\nu_{1}(t)B_{1}, \non \\
\epsilon_{N}&=&\hbar \om a_{N}^{\dagger}a_{N} 
+\frac{1}{2}\hbar\mu_{N-1}(t)A_{N-1} +\frac{1}{2}\hbar\nu_{N-1}(t)B_{N-1}\non \\
\end{eqnarray}

Note that the number density operator is independent of the
noise.  The energy density, however, has both a noise independent component
(which is proportional to number density) and a noisy component linear
in coefficients $\mu_i(t)$ or $\nu_i(t)$. Thus, we will need to take
both types of averages over noise discussed in Sec. \ref{2}.  We have
for $\overline{\bra n_i(t)\ket}$,
\begin{eqnarray}
\overline{\bra n_i(t)\ket}={\rm Tr} (n_i e^{-i{\cal L}_0t-{\cal
M}t}\rho(0)). \label{avenum}
\end{eqnarray}
Following procedure in Sec. \ref{2} and using the explicit form of the
second moments (Eq. \ref{corr}) we obtain,
\begin{eqnarray}
M\rho=\frac{\alpha^2_1}{2}\sum_i\left[A_i,\left[A_i,\rho\right]\right]
+\frac{\alpha^2_2}{2}\sum_i\left[B_i,\left[B_i,\rho\right]\right].\label{superM}
\end{eqnarray}
Here the summations extend from $1$ to $N$ for cyclic boundaries and
 from $1$ to $N-1$ for open-end boundaries.  In the case of
 $\overline{\bra \epsilon_i(t)\ket}$, we can evaluate the
 noise-dependent part using Eq. \ref{ave2}. Substituting $A_i$ or
 $B_i$ for $W$ and using Eq. \ref{corr} we obtain
\begin{eqnarray}
\overline{\bra \mu_i(t)A_i(t)\ket}=\overline{\bra \nu_i(t)B_i(t)\ket}=0.
\end{eqnarray}
Thus, for the special form of second moments in Eq. (\ref{corr}), time
 dependent part of $\epsilon_i$ does not contribute to $\overline{\bra
 \epsilon_i(t)\ket}$ and we have
\begin{eqnarray}
\overline{\bra \epsilon_i(t)\ket}=\hbar\omega\overline{\bra n_i(t)\ket}.
\end{eqnarray}
In order to arrive at equations of motion for $\overline{\bra
\epsilon_i(t)\ket}$ and $\overline{\bra n_i(t)\ket}$, we differentiate
both sides of Eq. (\ref{avenum}) with respect to time and obtain
\begin{eqnarray}
\frac{\partial \overline{\bra n_i(t)\ket}}{\partial t}&=&-{\rm Tr}(
n_i (i{\cal L}_0+{\cal M})e^{-i{\cal L}_0t-{\cal M}t}\rho(0) )
\nonumber \\ &=&-{\rm Tr}(\overline{\rho(t)}{\cal M}n_i). \label{ntd}
\end{eqnarray}
To obtain the last expression we used explicit form of superoperators
${\cal L}_0$ and ${\cal M}$, cyclic permutations under trace and the
fact that ${\cal L}_0 n_i =0$.
Using the explicit form of superoperator ${\cal M}$ we obtain the
following equations for $\overline{\bra n_i(t)\ket}$ and
$\overline{\bra \epsilon_i(t)\ket}$
\begin{eqnarray}
\frac{\partial\overline{\bra n_i(t)\ket}}{\partial
t}&=&\alpha^2\Delta_i \overline {\bra n_i(t)\ket} \label{Fick1} \\
\frac{\partial\overline{\bra \epsilon_i(t)\ket}}{\partial
t}&=&\alpha^2\Delta_i \overline{\bra \epsilon_i(t)\ket}
 \label{Fick}
\end{eqnarray}
here $\alpha^2=\alpha_1^2+\alpha_2^2$ and $\Delta_i$ is a discrete Laplacian whose action on some function
$f_i$ defined over discrete set of points $i$ is given by
\begin{eqnarray}
\Delta_if_i=-2f_i+f_{i+1}+f_{i-1}.
\end{eqnarray} 
In the case of cyclic boundary conditions Eqs. \ref{Fick1} and \ref{Fick} are valid
for all $i$'s, whereas for open end boundaries they are valid for all
$i$'s except $i=1$ and $i=N$ for which we have
\begin{eqnarray}
\frac{\partial\overline{\bra n_1(t)\ket}}{\partial
t}&=&\alpha^2(\overline{\bra n_2(t)\ket}-\overline{\bra
n_1(t)\ket}),\non \\ \frac{\partial\overline{\bra
n_N(t)\ket}}{\partial t}&=&\alpha^2(\overline{\bra
n_{N-1}(t)\ket}-\overline{\bra n_N(t)\ket}),\non \\
\frac{\partial\overline{\bra \epsilon_1(t)\ket}}{\partial
t}&=&\alpha^2(\overline{\bra \epsilon_2(t)\ket}-\overline{\bra
\epsilon_1(t)\ket}), \non\\ \frac{\partial\overline{\bra
\epsilon_N(t)\ket}}{\partial t}&=&\alpha^2(\overline{\bra
\epsilon_{N-1}(t)\ket}-\overline{\bra \epsilon_N(t)\ket})
\end{eqnarray}
$\Delta_i$ is the discrete analogue of the continuous operator
$l^2\partial^2/ \partial x^2$ where $l$ is interatomic distance.  We
can see that for cyclic boundary conditions, $\overline{\bra
n_i(t)\ket}$ and $\overline{\bra \epsilon_i(t)\ket}$ satisfy discrete
versions of the diffusion and heat equations, respectively.  For open
end boundary conditions these equations are satisfied everywhere
except the boundaries.  In both cases the stationary solutions
correspond to constant $\overline{\bra n_i(t)\ket}$ and
$\overline{\bra \epsilon_i(t)\ket}$.  Coefficients of diffusion $K_d$
and energy density diffusion  $K_h$ are given by
\begin{eqnarray}
K_d=K_h=\alpha^2l^2 \label{coefficients}
\end{eqnarray}
In the case of diffusion coefficient $K_d$, Eq. (\ref{coefficients})
agrees with the earlier results \cite{ovchinnikov:733, madrukar1424}.
Time dependent diffusion coefficient obtained in  Ref. \cite{madrukar1424}
reduces to our result Eq. (\ref{coefficients}) for the special case of Hamiltonian
(\ref{Ham}) because all time dependent terms vanish for Hamiltonians with
zero averaged hopping coefficient.
We also note here that the closed equations for $\overline{\bra
n_i(t)\ket}$ and $\overline{\bra \epsilon_i(t)\ket}$ are due to the
simple form of Hamiltonian
(\ref{Ham}) and the resulting
structure of superoperator ${\cal M}$.  For more general Hamiltonians, such as those
considered in Refs. \cite{ovchinnikov:733, madrukar1424}, closed
equations exist only for the noise-averaged single particle density
matrix, $\overline{\bra a_i^{\dagger}a_j\ket}$. However, such more general models
do not conserve noise averaged energy of the system and therefore cannot be used 
for studies of heat conductivity which is the primary focus of  this paper.
Including on-site noise terms in Hamiltonian (\ref{Ham}) does 
conserve the noise averaged energy. However, inclusion of these terms 
does not affect the transport equations given by 
Eqs. (\ref{Fick1}, \ref{Fick}).

\section{Coupling to the reservoirs}

We would like to verify if a linear energy density distribution will
be achieved if the lattice is placed between two stochastic
Langevin-type reservoirs.
To model coupling of the lattice to such reservoirs we will use our
original open-end lattice model and modify the equations of motions
for the operators for the first and last sites.  These modifications
are achieved by, firstly, including additional time-dependent noise
terms of the form $\hbar\eta_1(t)(a_1^{\dagger}+a_1)$ and
$\hbar\eta_N(t)(a_N^{\dagger}+a_N)$ for the first and last sites to
Hamiltonian (\ref{Ham}) and, secondly, by introducing a
non-Hamiltonian damping for the creation and annihilation operators at
the terminal sites with decay rates of, respectively, $\gamma_1$ and
$\gamma_N$.  The noise sources $\eta_1(t)$ and $\eta_N(t)$ are assumed
to be not correlated to other noises in the Hamiltonian (\ref{Ham}) or
to each other and hence can be treated on the same basis as the
hopping integral noises. We assume that
\begin{eqnarray}
\overline{\eta_1(t)\eta_1(t')}=\xi_1^2\delta(t-t'),
\,\,\overline{\eta_N(t)\eta_N(t')}=\xi_N^2\delta(t-t').
\end{eqnarray}

Note that the stochastic coupling that we consider corresponds to
reservoirs that exchange both energy and particles with the
lattice. In the case of bosons, modified equations for energy density
of the first and last sites are given by
\begin{eqnarray}
\frac{\partial\overline{\bra \epsilon_1\ket}}{\partial t}&=&
-(2\gamma_1 +\alpha^2)\overline{\bra
  \epsilon_1\ket}+\alpha^2\overline{\bra \epsilon_2\ket} +\om\xi_1^2
\\ \frac{\partial\overline{\bra \epsilon_N\ket}}{\partial t}&=&
-(2\gamma_N +\alpha^2)\overline{\bra
  \epsilon_N\ket}+\alpha^2\overline{\bra \epsilon_{N-1}\ket}
+\om\xi_N^2
\end{eqnarray}
Equations of motion for energy densities at other sites remain
unchanged and are given in Eq. (\ref{Fick}).  The stationary solutions
of equations correspond to a linear distribution of $\overline{\bra
\epsilon_i\ket}$.  The energy gradient $\theta$
\begin{eqnarray}
\theta=\frac{\overline{\bra \epsilon_N\ket}-\overline{\bra
\epsilon_1\ket}}{N-1},
\end{eqnarray}
is given in terms of reservoir coupling parameters by
\begin{eqnarray}
\theta=\frac{\om(\gamma_1\xi_N^2-\gamma_2\xi_N^2)}
{2(N-1)\gamma_1\gamma_N+\alpha^2(\gamma_1+\gamma_N)}
\end{eqnarray}
while for the stationary state
\begin{eqnarray}
\overline{\bra
\epsilon_1\ket}=\frac{\om\xi_1^2+\alpha^2\theta}{2\gamma_1}, \qquad
\overline{\bra
\epsilon_N\ket}=\frac{\om\xi_N^2-\alpha^2\theta}{2\gamma_1}.
\end{eqnarray}
In the case of fermionic or spin operators, equations of motion for
$\overline{\bra \epsilon_1\ket}$ and $\overline{\bra \epsilon_N\ket}$
are
\begin{eqnarray}
\frac{\partial\overline{\bra \epsilon_1\ket}}{\partial t}&=&
-(2\gamma_1+\alpha^2+2\omega\xi_1^2)\overline{\bra
  \epsilon_1\ket}+\alpha^2\overline{\bra \epsilon_2\ket} +\om\xi_1^2
\non \\ \frac{\partial\overline{\bra \epsilon_N\ket}}{\partial t}&=&
-(2\gamma_N +\alpha^2+2\omega\xi_N^2)\overline{\bra
  \epsilon_N\ket}+\alpha^2\overline{\bra \epsilon_{N-1}\ket}
+\om\xi_N^2. \non \\
\end{eqnarray}
Again, the stationary state corresponds to the linear distribution of
$\overline{\bra \epsilon_i\ket}$ with the energy density 
gradient  given by
\begin{widetext}
\begin{eqnarray}
\theta=\frac{\om(\gamma_1\xi_N^2-\gamma_N\xi_N^2)}
{2(N-1)(\gamma_1-\omega\xi_1^2)(-\gamma_N+\omega\xi_N^2)
+\alpha^2(-\gamma_1-\gamma_N+\omega \xi_1^2+\omega\xi_2^2)}
\end{eqnarray}
and
\begin{eqnarray}
\overline{\bra
\epsilon_1\ket}=\frac{\om\xi_1^2+\alpha^2\theta}{2(\gamma_1+\omega\xi_1^2)},
\qquad \overline{\bra
\epsilon_N\ket}=\frac{\om\xi_N^2-\alpha^2\theta}{2(\gamma_N+\omega\xi_N^2)}.
\end{eqnarray}
\end{widetext}
Since in this case we are dealing with non-equlilibrium situation 
there is no unique way of introducing temperature in this case.  However, we can 
say that the local temperature is proportional to the average energy density at a given site.
Hence, we recover the macroscopic heat equation starting from a microscopic quantum 
description.

\section{Discussion}

In this paper, we developed a quantum mechanical model for energy and
particle transport in a linear chain with stochastic hopping
coefficients.  We present a set of specific conditions in which the
noise-averaged energy expectation value will remain constant.  Under
such conditions we arrive at a quantum mechanical law for heat
transport.  Furthermore, we consider the case in which the chain can
exchange both particles and energy with reservoirs on either end.  
Both the energy density gradient and particle
density gradient can be obtained in closed form.  We consider only
gaussian noise for our stochastic coefficients.  More detailed
analysis will be required to develop a complete understanding of how 
well the averages represent the behavior of individual
members of the ensemble.  To reconfirm our theoretical results we performed a series 
of numerical simulations for our model (\ref{Ham}) with a particle placed at one of the sites
at $t=0$. As expected normal diffusion was observed for noise averaged time evolution for wide
range of noise strengths. 
The simplicity of this model allows
us to develop a more general treatment of quantum transport in a wide
range of physical systems ranging from resonant electronic energy
transport in DNA chains and organic glasses to Joule heating in molecular wires.

\begin{acknowledgments}
This work was funded in part through grants from the National Science
Foundation and the Robert A. Welch foundation.
\end{acknowledgments}

\end{document}